\def\draftversion{false}
\RequirePackage{ifthen}
\ifthenelse{\equal{\draftversion}{true}}{
  \documentclass[galley,aps,pra,10pt,amsmath,amssymb,
    superscriptaddress,nofootinbib,longbibliography]{revtex4-2}
}{
  \documentclass[twocolumn,aps,pra,10pt,amsmath,amssymb,
    superscriptaddress,nofootinbib,longbibliography]{revtex4-2}
}
\usepackage{graphicx}
\usepackage[usenames,dvipsnames]{color} 
\usepackage{bm} 
\usepackage{soul} 
\usepackage{mathtools}
\usepackage{tabularx}
\usepackage{dcolumn}
\usepackage{bbold}
\usepackage{hyperref}

\usepackage[normalem]{ulem}
\newcommand{\stkout}[1]{\ifmmode\text{\sout{\ensuremath{#1}}}\else\sout{#1}\fi}

\hypersetup{
    colorlinks=true,       
    linkcolor=blue,          
    citecolor=blue,       
    filecolor=magenta,      
    urlcolor=blue          
}

\ifthenelse{\equal{\draftversion}{true}}{
  \marginparwidth 2.7in
  \marginparsep 0.5in
  \newcounter{comm} 
  \def\commnext{\stepcounter{comm}}
  \def\commtext{{\bf\color{blue}[\arabic{comm}]}}
  \def\commmar{{\bf\color{blue}[\arabic{comm}]}}
  \def\dvm#1{\commnext\marginpar{\small DV\commmar: #1}\commtext}
  \def\tcm#1{\commnext\marginpar{\small TC\commmar: #1}\commtext}

}{
  \def\dvm#1{}
  \def\tcm#1{}

}

\newcommand{\beq}{\begin{equation}}
\newcommand{\eeq}{\end{equation}}
\newcommand{\bea}{\begin{eqnarray}}
\newcommand{\eea}{\end{eqnarray}}

\newcommand{\eq}[1]{Eq.~(\ref{eq:#1})}

\newcommand{\eqs}[2]{Eqs.~(\ref{eq:#1}) and (\ref{eq:#2})}

\newcommand{\ket}[1]{\vert#1\rangle}

\begin{document}

\title{ 
Exact expression for the Berry connection in the projection gauge
}

\author{Trey Cole}
\affiliation{
Department of Physics \& Astronomy, Rutgers University,
Piscataway, New Jersey 08854, USA}

\author{David Vanderbilt}
\affiliation{
Department of Physics \& Astronomy, Rutgers University,
Piscataway, New Jersey 08854, USA}

\begin{abstract}
The Berry connection encodes the momentum-space geometry of occupied Bloch states in gapped insulators and plays a central role in topological materials. While gauge-invariant quantities can be evaluated from overlap matrices between neighboring $k$ points, accessing the Berry connection itself as a smooth field requires specifying a continuous gauge over the Brillouin zone. Wannier-based workflows achieve this through projection onto localized orbitals, enabling stable evaluation of geometric quantities and response functions. In this setting, the Berry connection enters directly in Wannier-interpolated calculations of polarization, Berry curvature, and related response functions. In practical implementations, however, the projection-gauge Berry connection is typically constructed from finite-difference overlaps between neighboring $k$ points, discretizing momentum derivatives and introducing errors tied to $k$-mesh spacing and gauge alignment. These effects can become numerically delicate in systems with small band gaps or when evaluating higher-order responses such as the Chern–Simons axion angle. Here, we derive an exact expression for the non-Abelian Berry connection in the projection gauge that is local in crystal momentum. Starting from projected and orthonormalized Bloch-like states, we obtain a closed-form equation expressed entirely in terms of $k$-local quantities. We validate the formulation in one and three dimensions by computing the Berry phase and Chern–Simons axion angle in tight-binding models. The resulting framework provides a stable route to evaluating geometric properties within Wannier interpolation schemes and future first-principles implementations.
\end{abstract}

\maketitle

\section{Introduction}

The modern theory of electronic structure has revealed that many physical properties of crystalline solids are governed not only by band energies, but by the geometry of the occupied Bloch states in momentum space. This geometry is encoded in the Berry connection and its derivatives, which define a gauge structure over the Brillouin zone. Observable consequences of this structure include electric polarization, anomalous and nonlinear Hall effects, orbital magnetization, optical and transport responses, and the magnetoelectric coupling \cite{xiao2010}. These quantities depend on the geometric structure of the occupied-band bundle, rather than solely on the dispersion relation of the bands. In multiband systems, the relevant geometric object is the non-Abelian Berry connection, 
\begin{equation}
    \label{eq:berry_connection}
    \left( \mathcal{A}_\mu \right)_{mn}
    =
    i \langle u_{m\mathbf{k}} | \partial_{\mu} u_{n\mathbf{k}} \rangle ,
\end{equation}
where $|u_{n\mathbf{k}}\rangle$ are the occupied cell-periodic Bloch eigenstates and $\partial_{\mu}$ denotes differentiation with respect to the $\mu$-th Cartesian component of crystal momentum \cite{Vanderbilt_2018}.

In practical electronic-structure calculations, the Bloch eigenstates are obtained by diagonalizing the Hamiltonian independently at each $k$ point. The resulting gauge, here referred to as the \emph{Hamiltonian gauge}, is therefore arbitrary up to $k$-dependent single-band phase factors and may have nonanalytic behavior at degeneracies in the Brillouin zone. As a consequence, direct finite-difference evaluations of the Berry connection in this gauge are numerically unstable and ill-defined, even though gauge-invariant quantities constructed from them may remain well behaved.

Evaluating $\mathcal{A}_\mu$ itself requires specifying a smooth and continuous gauge over the Brillouin zone. The Berry connection is not gauge-covariant and, under a smooth and continuous $k$-dependent unitary rotation $U(\mathbf{k})$ within the occupied subspace, transforms as
\begin{equation}
\label{eq:tildeA}
    \tilde{\mathcal{A}}_\mu
    =
    U^\dagger \mathcal{A}_\mu U
    + i U^\dagger \partial_{\mu} U .
\end{equation}
For any initially discontinuous gauge, the unitary that smoothens the gauge will also itself be discontinuous, making $\partial_{\mu} U$ ill-defined. 

Wannier-based workflows provide a practical solution by constructing a smooth gauge through projection onto localized trial orbitals, optionally followed by disentanglement and maximal localization \cite{marzari1997, souza2001, marzari2012}. The resulting \emph{projection (or Wannier) gauge} enables stable evaluation of geometric quantities. In this setting, the projection-gauge Berry connection encodes the off-diagonal position matrix elements in the Wannier basis \cite{xiao2010, Vanderbilt_2018}. These terms represent what is often referred to as the ``external" contributions to the band geometry, in contrast to the ``internal" contributions that are captured by the tight-binding eigenvectors. These external terms are needed, for example, when computing electric-dipole transition moments for the interband optical conductivity \cite{yates2007}, and Wannier-interpolating the Berry curvature \cite{wang2006}. 

However, in standard Wannier codes, such as \textsc{Wannier90} \cite{MOSTOFI20142309}, the projection-gauge Berry connection is not computed analytically; instead, it is approximated by finite-difference overlaps between the first-principles Bloch eigenstates at neighboring $k$-points \cite{marazzo2024, marzari1997}. Higher-order generalizations of this finite-difference scheme for computing $\tilde{A}_\mu$ have recently been introduced \cite{lihm2026}. These methods are robust and practical, and have been central to the success of Wannier interpolation across a wide range of applications. At the same time, because momentum derivatives are discretized from the outset, their accuracy is ultimately controlled by the $k$-mesh spacing, the smoothness of the chosen gauge, and the quality of the gauge alignment between neighboring points. These effects may lead to appreciable quantitative differences in derived observables, including optical conductivities, in certain regimes \cite{thummler2026}.

In this work, we derive an explicit closed-form expression for the non-Abelian Berry connection directly in the simple projection gauge (i.e., obtained as in the usual Wannierization procedure but without maximal localization) without recourse to finite differences or neighboring-$k$ overlaps. Starting from the projected and orthonormalized Bloch-like states produced by the projection method, we obtain a single-point, Kubo-like formula for $\tilde{A}_\mu(\mathbf{k})$ expressed entirely in terms of overlap matrices, position matrix elements of the trial orbitals, and interband velocity matrix elements. The connection is thus determined exactly at each $k$ from local information in momentum space. 

We apply the formulation derived in Sec.~\ref{sec:conn} to tight-binding models in one and three dimensions. In Sec.~\ref{sec:berry_phase_proj}, we compute the Berry phase of the Su--Schrieffer--Heeger model, providing a controlled one-dimensional test of the single-point construction. In Sec.~\ref{sec:FKM}, we evaluate the Chern--Simons axion angle in the Fu--Kane--Mele model, demonstrating that the analytic projection-gauge connection remains stable in higher-dimensional, non-Abelian response calculations. Together, these tests illustrate the numerical robustness of the approach.

\section{Projection gauge}
\label{sec:proj_gauge}

A central practical difficulty in evaluating Berry connections is that the eigenstates produced by a band-structure calculation are defined only up to an arbitrary $k$-dependent phase rotation acting on each individual band. As a result, the ``Hamiltonian gauge'' obtained by diagonalizing the Hamiltonian independently at each $k$ typically exhibits random phases and nonanalytic behavior across the Brillouin zone. The projection method provides a simple and widely used way to impose a smooth periodic choice of frame for the occupied subspace by aligning it to a set of localized trial orbitals and orthonormalizing the result \cite{souza2001, marzari2012}.

Let $\{|\psi_{n\mathbf{k}}\rangle\}$ denote the Bloch eigenstates of the Hamiltonian at crystal momentum $\mathbf{k}$, with band index $n$ and $\{|u_{n\mathbf{k}}\rangle\}$ their cell-periodic counterparts. We assume a gapped insulator with $J$ occupied bands and denote the occupied-band projector by
\begin{equation}
\label{eq:Pocc}
\hat{P}(\mathbf{k}) \equiv \sum_{n}^{\mathrm{occ}} |u_{n\mathbf{k}}\rangle\langle u_{n\mathbf{k}}|.
\end{equation}
The projection method begins by choosing a set of $J$ localized trial orbitals $\{|g_m\rangle\}_{m=1}^J$ in the home unit cell.
Physically, these are intended to resemble the desired Wannier functions (or, more generally, to span a subspace close to the occupied manifold).

Each $|g_m\rangle$ has a phase-twisted counterpart
\begin{equation}
    |t_{m\mathbf{k}} \rangle = e^{-i\mathbf{k}\cdot \mathbf{r}} | g_m \rangle
\end{equation}
that can be projected onto the occupied subspace at $\mathbf{k}$ to define a set of ``Bloch-like'' states
\begin{equation}
\label{eq:bloch-like}
|\phi_{m\mathbf{k}}\rangle
\equiv
\hat{P}(\mathbf{k})|t_{m\mathbf{k}}\rangle
=
\sum_{n}^{\mathrm{occ}} |u_{n\mathbf{k}}\rangle\langle u_{n\mathbf{k}}|t_{m \mathbf{k}}\rangle.
\end{equation}
These states lie entirely in the occupied subspace by construction, but they are generally neither orthonormal nor linearly independent.

Because the projected states in \eq{bloch-like} are not orthonormal, we introduce their Gram matrix
\begin{equation}
\label{eq:O_def}
(\mathcal{O}_{\mathbf{k}})_{mn}
\equiv
\langle \phi_{m\mathbf{k}}|\phi_{n\mathbf{k}}\rangle.
\end{equation}
It is convenient to collect the overlaps between occupied Bloch states and trial orbitals into the $J\times J$ matrix
\begin{equation}
\label{eq:ovlp}
S_{nm}(\mathbf{k})
\equiv
\langle u_{n\mathbf{k}}|t_{m \mathbf{k}}\rangle,
\qquad n\in\mathrm{occ}.
\end{equation}
Using \eqs{bloch-like}{ovlp} and the orthonormality of the Bloch eigenstates, one finds immediately that
\begin{equation}
\label{eq:O_SdagS}
\mathcal{O}_{\mathbf{k}} = S^\dagger_{\mathbf{k}} S_{\mathbf{k}}.
\end{equation}
Thus $\mathcal{O}_{\mathbf{k}}$ is Hermitian and positive semidefinite.
A necessary condition for the projection method to define a $J$-dimensional frame is that $\mathcal{O}_{\mathbf{k}}$ be invertible,
i.e., that the projected states $\{|\phi_{m\mathbf{k}}\rangle\}$ be linearly independent. In practice, small eigenvalues of $\mathcal{O}_{\mathbf{k}}$ signal a poorly conditioned projection and can lead to numerical instability.

When $\mathcal{O}_{\mathbf{k}}$ is invertible, we obtain an orthonormal frame by symmetric (L\"owdin) orthonormalization,
\begin{equation}
\label{eq:tildepsi_from_phi}
|\tilde{u}_{n\mathbf{k}}\rangle
\equiv
\sum_m |\phi_{m\mathbf{k}}\rangle\,(\mathcal{O}^{-1/2}_{\mathbf{k}})_{mn}.
\end{equation}
By construction,
\begin{equation}
\langle \tilde{u}_{m\mathbf{k}}|\tilde{u}_{n\mathbf{k}}\rangle
=
[(\mathcal{O}^{-1/2}_{\mathbf{k}})^\dagger \mathcal{O}_{\mathbf{k}} (\mathcal{O}^{-1/2}_{\mathbf{k}})]_{mn}
=
\delta_{mn},
\end{equation}
so $\{|\tilde{u}_{n\mathbf{k}}\rangle\}$ acts as an orthonormal basis of the occupied subspace at each $\mathbf{k}$.
We refer to this orthonormal frame as the \emph{projection gauge} (or equivalently, the Wannier gauge produced by projection).

For an isolated group of bands in an ordinary insulator, $\hat P(\mathbf{k})$ can be chosen smooth over the Brillouin zone, and for suitably chosen localized trial orbitals the matrices $S(\mathbf{k})$ and $\mathcal{O}(\mathbf{k})$ vary smoothly as well,
leading to a smooth and (when topologically allowed) periodic frame $\{|\tilde{\psi}_{n\mathbf{k}}\rangle\}$
\cite{marzari1997, marzari2012}. In $\mathbb{Z}_2$ topological insulators, a smooth and periodic \emph{time-reversal-symmetric} gauge is obstructed; nevertheless, a smooth periodic gauge can still be constructed by choosing trial orbitals that explicitly break $\mathcal{T}$, thereby avoiding the constraint that enforces the obstruction \cite{soluyanov2011}.

\section{Non-Abelian connection in the projection gauge}
\label{sec:conn}

The goal of this section is to obtain the non-Abelian Berry connection associated with the projection-gauge frame
$\{|\tilde{u}_{n\mathbf{k}}\rangle\}$ constructed in Sec.~\ref{sec:proj_gauge}. We will derive a closed, \emph{single-$k$} formula for the projection-gauge connection, expressed entirely in terms of
(i) overlaps to trial orbitals, (ii) position matrix elements involving the trial orbitals, and
(iii) interband (Kubo-like) velocity matrix elements.
No finite differences and no overlaps between neighboring $k$ points are required.

The projection-gauge connection is defined as
\begin{equation}
\label{eq:Atilde_def_repeat}
\tilde{\mathcal{A}}_{\mu,mn}
\equiv
i\langle \tilde{u}_{m\mathbf{k}}|\partial_{\mu}\tilde{u}_{n\mathbf{k}}\rangle .
\end{equation}
It is convenient to use matrix notation, in which $|\phi\rangle$ is the column vector of states $|\phi_{n\mathbf{k}}\rangle$,
and similarly for $|\tilde{u}\rangle$. We also drop $\mathbf{k}$ labels, which are now always implicit.
Then
\begin{equation}
\label{eq:tildepsi_matrix}
|\tilde{u}\rangle = |\phi\rangle\,\mathcal{O}^{-1/2}.
\end{equation}
Differentiating with respect to $k_\mu$ gives
\begin{equation}
\label{eq:dtildepsi_matrix}
\partial_\mu|\tilde{u}\rangle
=
(\partial_\mu|\phi\rangle)\,\mathcal{O}^{-1/2}
+
|\phi\rangle\,\partial_\mu(\mathcal{O}^{-1/2}).
\end{equation}
Substituting into \eqref{eq:Atilde_def_repeat} and using $\langle\phi|\phi\rangle=\mathcal{O}$ yields the identity
\begin{equation}
\label{eq:Atilde_split_repeat}
\tilde{\mathcal{A}}_\mu
=
i\,\mathcal{O}^{-1/2}\,\langle\phi|\partial_\mu\phi\rangle\,\mathcal{O}^{-1/2}
\;+\;
i\,\mathcal{O}^{1/2}\,\partial_\mu(\mathcal{O}^{-1/2}).
\end{equation}
Using $\partial_\mu(\mathcal{O}^{1/2}\mathcal{O}^{-1/2})=0$ we have $\mathcal{O}^{1/2}\,\partial_\mu(\mathcal{O}^{-1/2})=-(\partial_\mu \mathcal{O}^{1/2})\mathcal{O}^{-1/2}$, hence
\begin{equation}
\label{eq:Atilde_key_identity}
\tilde{\mathcal{A}}_\mu
=
i\,\mathcal{O}^{-1/2}\,\langle\phi|\partial_\mu\phi\rangle\,\mathcal{O}^{-1/2}
\;-\;
i\,\partial_\mu(\mathcal{O}^{1/2})\,\mathcal{O}^{-1/2}.
\end{equation}

Equation \eqref{eq:Atilde_key_identity} is at the core of the derivation. It shows that $\tilde{\mathcal{A}}_\mu$ is determined once we can compute (i) the derivative of $\mathcal{O}^{1/2}$, and (ii) the projected-state matrix element $\langle\phi|\partial_\mu\phi\rangle$, both at fixed $k$.

\subsection{Sylvester equation for $\partial_\mu \mathcal{O}^{1/2}$}

We next obtain $\partial_\mu \mathcal{O}^{1/2}$ in closed form. Differentiating the identity $(\mathcal{O}^{1/2})^2=\mathcal{O}$ gives
\begin{equation}
\label{eq:sylvester_P}
\mathcal{O}^{1/2}(\partial_\mu \mathcal{O}^{1/2})+(\partial_\mu \mathcal{O}^{1/2})\mathcal{O}^{1/2}=\partial_\mu \mathcal{O},
\end{equation}
which is a Sylvester equation of the standard form $MX + XM = Y$ for the unknown $X=\partial_\mu \mathcal{O}^{1/2}$, with $M=\mathcal{O}^{1/2}$ and $Y=\partial_\mu \mathcal{O}$. 

Consider the Sylvester map $\mathcal{L}_M:\,X\mapsto MX+XM$ for a Hermitian positive semidefinite matrix $M$. If $M$ is positive definite, $\mathcal{L}_M$ is invertible on the full matrix space. More generally, when $M$ has a nontrivial kernel, $\mathcal{L}_M$ remains invertible on the subspace of matrices with no components in $\ker{M} \otimes \ker{M}$. On this subspace, the Sylvester equation admits a unique solution, which we denote $X=\mathcal{L}_M^{-1}[Y]$. 

With this notation, the formal solution to \eq{sylvester_P} reads
\begin{equation}
\label{eq:sylv}
\partial_\mu \mathcal{O}^{1/2}
=
\mathcal{L}_{\mathcal{O}^{1/2}}^{-1}\!\left[\partial_\mu \mathcal{O}\right].
\end{equation}
To evaluate \eq{sylv} numerically, we transform \eq{sylvester_P} into the eigenbasis of $\mathcal{O}^{1/2}$, letting
\begin{equation}
\label{eq:M_diag}
\mathcal{O}^{1/2} = V\Sigma V^\dagger,
\;\;
\Sigma=\mathrm{diag}(\sigma_1,\dots,\sigma_J),
\end{equation}
where $V$ is unitary and $\sigma_i \ge 0$ are the eigenvalues\footnote{Equivalently, $\sigma_i$ are the singular values of the projection matrix $S$.} of $\mathcal{O}^{1/2}$. Some algebraic manipulation yields the componentwise solution (see Appendix \ref{apdx:sylvester} for the full derivation),
\begin{equation}
\label{eq:syl_sol}
\mathcal{L}_{\mathcal{O}^{1/2}}^{-1}[\partial_\mu \mathcal{O}]_{mn}
=
V_{mi}\left[
\frac{(V^\dagger (\partial_\mu \mathcal{O}) V)_{ij}}{\sigma_i+\sigma_j}
\right]V^\dagger_{jn} ,
\end{equation}
where the division is understood to be elementwise inside the brackets, and outside the brackets, repeated indices are implicitly summed over.

Because $\sigma_i \ge 0$, the only potentially vanishing denominators satisfy $\sigma_i+\sigma_j=0$, which implies $\sigma_i=\sigma_j=0$. Therefore, $\mathcal{L}_{\mathcal{O}^{1/2}}^{-1}$ is well defined provided that $\partial_\mu \mathcal{O}$ has no components in the null--null block $\ker(\mathcal{O}^{1/2}) \otimes \ker(\mathcal{O}^{1/2})$. If $\mathcal{O}^{1/2}$ is strictly positive definite (all $\sigma_i>0$), $\mathcal{L}_{\mathcal{O}^{1/2}}$ is invertible on the full matrix space.

In practice, very small $\sigma_i$ correspond to trial-orbitals with weak support on the occupied subspace, leading to an ill-conditioned projection and numerical instability in \eq{syl_sol}. This can be mitigated by improving the choice of trial orbitals or, when appropriate, restricting to the well-conditioned subspace. If the projection loses rank---for example, due to a topological obstruction---some $\sigma_i$ vanish at certain $k$, signaling that a globally smooth projection gauge cannot be constructed. 

The remaining ingredient is the derivative of the overlap matrix $\mathcal{O}_{mn}=\langle\phi_m|\phi_n\rangle$. By direct differentiation,
\begin{equation}
\label{eq:parmu_O}
\partial_\mu \mathcal{O}
=
\langle \partial_\mu \phi|\phi\rangle
+
\langle \phi|\partial_\mu \phi\rangle
=
\langle \phi|\partial_\mu \phi\rangle + \text{h.c.},
\end{equation}
where ``h.c.'' denotes the Hermitian conjugate of the matrix
$\langle \phi|\partial_\mu \phi\rangle$.
Notably, the same matrix $\langle \phi|\partial_\mu \phi\rangle$
also appears in the first term of Eq.~\eqref{eq:Atilde_key_identity}.
A closed, single-$k$ expression for this object will be derived in Sec.~\ref{sec:eval}.

\subsection{Single-$k$ evaluation of $\langle\phi|\partial_\mu\phi\rangle$}
\label{sec:eval}

We now turn our attention to the evaluation of $\langle\phi|\partial_\mu\phi\rangle$, which is the central quantity that appears explicitly in the first term of \eq{Atilde_key_identity} and implicitly in the second term via \eq{sylv}. To show that it can be expressed only in terms of single-$k$ ingredients, we differentiate \eq{bloch-like} to get
\begin{equation}
\label{eq:phidphi_split}
\langle\phi_m|\partial_\mu\phi_n\rangle
=
\langle t_m|\hat P\,(\partial_\mu \hat P)\,|t_n\rangle
+
\langle t_m|\hat P\,\partial_\mu|t_n\rangle
\end{equation}
which cleanly separates two physical effects: (i) the explicit $k$-dependence of the $k$-periodic trial orbital, which produces a term proportional to the position operator, and (ii) the $k$-dependence of the occupied projector $\hat P(\mathbf{k})$, which encodes interband mixing and produces a Kubo-like contribution. We evaluate these two terms in turn.

\paragraph*{(i) Trial-orbital (position-matrix-element) term.}
Using $|t_n\rangle=e^{-i\mathbf{k}\cdot\hat{\mathbf r}}|g_n\rangle$ one finds
\begin{equation}
i\,\partial_\mu |t_n\rangle = \hat r_\mu |t_n\rangle .
\end{equation}
It is therefore natural to define the occupied--trial position matrix elements
\begin{equation}
\label{eq:Rmu_def_here}
(\mathcal{R}_\mu)_{mn}
\equiv
\langle u_{m}|\hat r_\mu|t_{n}\rangle,
\qquad m\in\mathrm{occ},
\end{equation}
Then, returning to the second term in \eq{phidphi_split},
\begin{align}
i\langle t_m|\hat P\,\partial_\mu|t_n\rangle
&=
\sum_{\ell}^{\mathrm{occ}}
\langle t_m|u_\ell\rangle\,\langle u_\ell|\hat r_\mu|t_n\rangle
\nonumber\\
&=
\sum_{\ell}^{\mathrm{occ}} S^*_{\ell m}\,(\mathcal{R}_\mu)_{\ell n}
=
\big(S^\dagger \mathcal{R}_\mu\big)_{mn}.
\label{eq:trial_term_result}
\end{align}
Thus the trial-orbital contribution to $i\langle\phi|\partial_\mu\phi\rangle$ is simply $S^\dagger \mathcal{R}_\mu$.

\paragraph*{(ii) Interband (Kubo-like) projector-derivative term.}
The remaining term involves $\hat P(\partial_\mu\hat P)$, which only has support between occupied and unoccupied spaces because
$\hat P(\partial_\mu\hat P)\hat P=0$.
Introducing the complementary projector $\hat Q=1-\hat P$ and inserting a complete set of unoccupied eigenstates, one finds
\begin{equation}
\label{eq:PdP_expand}
\hat P(\partial_\mu \hat P)
=
\hat P(\partial_\mu\hat P)\hat Q
=
\sum_{\ell}^{\mathrm{occ}}\sum_{c}^{\mathrm{empty}}
|u_\ell\rangle\langle \partial_\mu u_\ell| u_c\rangle\langle u_c|.
\end{equation}
The cross-gap matrix elements of the non-Abelian Berry connection matrix in the Hamiltonian gauge can be written as
\begin{equation}
\label{eq:Vmu_def_here}
(\mathcal{D}^{H}_\mu)_{\ell c}
= i\langle u_\ell | \partial_\mu u_c \rangle = i
\frac{\langle u_{\ell }|\partial_{\mu}H|u_{c}\rangle}{E_{c}-E_{\ell}}.
\;\; \ell \in\mathrm{occ},\ c\in\mathrm{empty}.
\end{equation}
It is also convenient to define the complementary overlap matrix
\begin{equation}
\label{eq:Sperp}
(S^\perp)_{c n}\equiv \langle u_{c}|t_{n}\rangle.
\qquad c\in\mathrm{empty}.
\end{equation}
Then
\begin{align}
i\langle t_m|\hat P(\partial_\mu\hat P)|t_n\rangle
&=
i\sum_{\ell}^{\mathrm{occ}}\sum_{c}^{\mathrm{empty}}
\langle t_m|u_\ell \rangle
\langle \partial_\mu u_\ell | u_c\rangle
\langle u_c|t_n\rangle
\nonumber\\
&= -
\sum_{\ell}^{\mathrm{occ}}\sum_{c}^{\mathrm{empty}}
S^*_{\ell m}\,
\mathcal{D}^H_{\mu,\ell c}
S^\perp_{c n}.
\label{eq:interband_term_mid}
\end{align}
This is precisely a Kubo-like expression that takes the compact matrix form
\begin{equation}
\label{eq:interband_term_result}
i\langle t|\hat P(\partial_\mu\hat P)|t\rangle
=
- S^\dagger\,\mathcal{D}^H_\mu\,S^\perp .
\end{equation}
Combining \eqref{eq:trial_term_result} and \eqref{eq:interband_term_result}, we obtain
\begin{equation}
\label{eq:phi_dphi_final}
i\langle\phi|\partial_\mu\phi\rangle
=
S^\dagger\!\left(\mathcal{R}_\mu - \mathcal{D}^H_\mu S^\perp\right).
\end{equation}
Equation \eqref{eq:phi_dphi_final} is the desired single-$k$ expression for the projected-state matrix element entering Eq.~\eqref{eq:Atilde_key_identity}.

\subsection{Final single-$k$ form}
Together with the Sylvester solution \eqs{sylv}{parmu_O}, \eq{phi_dphi_final} provides
a fully local construction of the projection-gauge Berry connection $\tilde{\mathcal{A}}_\mu(\mathbf{k})$ in terms of single-$k$ quantities. Substituting, we find
\begin{align}
\label{eq:tildeA_final}
    \tilde{\mathcal{A}}_\mu(\mathbf{k}) = \; &\mathcal{O}^{-1/2} S^\dagger \left( \mathcal{R}_\mu  -  \mathcal{D}^H_\mu S^\perp\right) \mathcal{O}^{-1/2} - \\ \nonumber &i \mathcal{L}^{-1}_{\mathcal{O}^{1/2}}[-i S^\dagger\left( \mathcal{R}_\mu  - \mathcal{D}^H_\mu S^\perp\right) + \mathrm{h.c.}]\;\mathcal{O}^{-1/2}
\end{align}
All dependence on $k$-derivatives has been eliminated in favor of the overlap matrices between Bloch eigenstates and trial orbitals, position matrix elements involving the trial orbitals, and a Kubo-like interband term that depends only on the occupied and empty projectors. As long as the occupied subspace is smooth and the projection is well-conditioned, this expression yields a well-defined and numerically stable non-Abelian connection without recourse to finite-difference schemes or overlaps between neighboring $k$ points.

We note that the present derivation is specific to the projection gauge, in which the gauge-defining states have an explicit closed-form dependence on $k$. For a gauge that incorporates additional numerical rotations---such as those arising from maximal localization or disentanglement---the Berry connection acquires a term $iU^\dagger \partial_\mu U$ whose closed-form evaluation is not accessible, since the unitary $U(k)$ is determined iteratively. It is possible to modify our approach to incorporate disentanglement without maximal localization, provided the isolated states of interest are included in the frozen energy window. However, for the isolated band manifolds on which the Berry connection is defined, disentanglement is unnecessary, and projection-only Wannierization is already well suited when the trial orbitals span the target space adequately.

In a first-principles setting, this same projection-gauge Berry connection would represent the ``external" contribution if computed from the projection matrices provided by \textsc{Wannier90}, for example. In this context, the sum over the unoccupied manifold can be avoided by iteratively solving the Sternheimer equation, which is a standard step in implementations of density-functional perturbation theory (DFPT). Here, the computational load, one linear-system solve per occupied band per Cartesian direction, is the same as that for a standard DFPT evaluation of the dielectric tensor ~\cite{Baroni1987, Gonze1995, Gonze1997, baroni2001}. In a tight-binding or Wannier-interpolated setting, the Hilbert space is finite-dimensional, and the sum over empty states in Eq.~(\ref{eq:Vmu_def_here}) is finite and exact.

Below, we test the applicability of the analytic expression for $\tilde{A}_\mu$ in a tight-binding setting for simplicity and clarity, allowing our expression to be compared against standard finite-difference constructions without additional complications. We form the projection gauge by projecting the tight-binding energy eigenstates onto a set of trial orbitals constructed as linear combinations of the tight-binding basis orbitals. We then compute the Berry connection in this gauge via Eq.~\ref{eq:tildeA_final}. All tight-binding calculations are performed using \textsc{PythTB} \cite{Cole_Python_Tight_Binding_2025, cole_2026_18727385}, which provides tools for constructing tight-binding models, generating smooth projection gauges, and evaluating Berry phases and related geometric quantities.

\section{Berry phase in the projection gauge}
\label{sec:berry_phase_proj}

We first test the single-$k$ expression for the Berry connection derived in Sec.~\ref{sec:conn} in the simplest possible setting: the Berry phase of a one-dimensional insulator. In one dimension, the Berry phase admits a well-known gauge-invariant formulation in terms of Wilson loops or overlap matrices between neighboring Bloch states. Our goal here is not to replace these established constructions, which are optimal in this setting, but rather to verify that the projection-gauge Berry connection reproduces the correct Berry phase in the continuum limit and to clarify the origin of discrepancies at finite $k$-mesh resolution.

\subsection{Berry phase in one dimension}

In one-dimensional insulating systems, the Berry phase accumulated by the occupied states upon traversing the Brillouin zone is a fundamental geometric quantity that admits a physical interpretation in terms of electronic polarization \cite{vanderbilt1993, resta1994}. We consider this as a first test case for our single-$k$ formulation of the Berry connection.

For a choice of Bloch states $\{|u_{n k}\rangle\}$ spanning the occupied subspace, the total Berry phase is given by
\begin{equation} 
\label{eq:berry_phase_1d}
\phi = \int_{\mathrm{BZ}} dk \, \text{Tr}\left[ \mathcal{A}(k) \right],
\end{equation}
where the trace is taken over the occupied bands. Because only the trace appears, the Berry phase depends only on the band-summed component of the non-Abelian connection. The Berry phase is invariant modulo $2\pi$ under smooth $k$-dependent gauge transformations of the occupied subspace.

In numerical calculations, the Berry phase is rarely evaluated by simple quadrature of the integral in Eq.~\eqref{eq:berry_phase_1d} on a $k$-mesh. Instead, it is typically obtained from the holonomy of the occupied subspace around the Brillouin zone, expressed in terms of Wilson loops or overlap matrices between neighboring $k$-points.
On a discrete $k$ mesh $\{k_j\}$, one defines the overlap (or link) matrices
\begin{equation}
\label{eq:overlap_matrix}
M_{mn}(k_j,k_{j+1}) = \langle u_{m k_j} | u_{n k_{j+1}} \rangle .
\end{equation}
which describes the parallel transport of the occupied subspace between neighboring $k$-points in momentum space. From these overlaps, one may define a discrete (band-traced) Berry connection associated with the link $(k_j, k_{j+1})$ as 
\begin{equation}
    \label{eq:discrete_connection}
    \mathrm{Tr}[\mathcal{A}^{\mathrm{FD}}(k_j)] = \frac{-1}{\delta k} \,\mathrm{Im} \, \ln \text{det} M(k_j, k_{j+1})
\end{equation}
where $\delta k  = k_{j+1} - k_j$ and the branch of the logarithm is chosen continuously along the path in $k$. For a single occupied band, this reduces to the familiar Abelian expression, while for multiple occupied bands, the determinant extracts the $U(1)$ part of the non-Abelian connection. 

The discrete Wilson loop (holonomy) associated with a closed traversal of the Brillouin zone is then defined as the ordered product
\begin{equation}
\label{eq:wilson_loop}
\mathcal{W} = M(k_0,k_1) M(k_1,k_2) \cdots M(k_{N-1},k_N),
\end{equation}
where $k_N = k_0 + G$ differs from $k_0$ by a reciprocal lattice vector. The Berry phase is obtained as,
\begin{equation}
\label{eq:berryphase_discretized}
\phi_{\mathrm{FD}}= - \mathrm{Im} \, \ln \text{det} \left[ \mathcal{W} \right]
   = \sum_j \mathrm{Tr}[\mathcal{A}^{\mathrm{FD}}(k_j)] \,\delta k .
\end{equation}

Alternatively, we can evaluate the Berry phase in our smooth projection gauge $\{|\tilde u_{n k}\rangle\}$, replacing $\mathcal{A}$ by $\tilde{\mathcal{A}}$ in Eq.~\eqref{eq:berry_phase_1d}. In view of the trace, we only need the total (band-summed) Berry connection
\begin{equation}
\mathrm{Tr}[\tilde{\mathcal{A}}^{\mathrm{SP}}(k)]
= \sum_n^{\mathrm{occ}} \tilde{\mathcal{A}}^{\mathrm{SP}}_{nn}(k) ,
\end{equation}
which can be evaluated at each $k$ point using the single-$k$ expressions derived in Sec.~\ref{sec:conn}, without finite differences in momentum space. The single-point Berry phase then becomes
\begin{equation}
\label{eq:riemann_sum}
\phi_{\text{SP}} = \sum_{j} \mathrm{Tr}[\tilde{\mathcal{A}}^{\mathrm{SP}}(k_j)]\,\delta k .
\end{equation}
By construction, $\phi_{\text{SP}}$ converges to the correct Berry phase in the continuum limit; however, its accuracy will depend on how well the $k$-mesh resolves variations in $\tilde{\mathcal{A}}(k)$.

\subsection{Results for tight-binding model}

To illustrate these considerations, we apply both approaches to the Su--Schrieffer--Heeger (SSH) model~\cite{ssh1979}, a one-dimensional dimerized tight-binding model consisting of alternating A and B sites located at $x=-a/4$ and $a/4$ respectively ($a$ is the lattice constant). In second-quantized form, the Hamiltonian is given by
\begin{align}
    \label{eq:ssh_ham}
    H = &\sum_{j} \big[v \, c^{\dagger}_{j,A} c_{j,B} + w\, c^{\dagger}_{j, B} c_{j+1, A} + \text{h.c.}\big] 
\end{align}
The model has inversion symmetry, which constrains the occupied-band Berry phase to be quantized to $0$ or $\pi$ (mod $2\pi$). Accordingly, the model exhibits two topologically distinct insulating phases with $\phi=0$ for $|v| > |w|$ and $\phi=\pi$ for $|v| < |w|$. The transition between phases occurs at $|v|=|w|$, where the bulk band gap closes. For $v=w$, the gap closes at the Brillouin-zone boundary $k$ = $\pi/a$, while for $v=-w$ it closes at the zone center $k=0$. We choose the parameterization $v=1$, $w=1.5$, corresponding to the phase with $\phi=\pi$ and a relatively large band gap. A simple trial orbital $|g\rangle = |A\rangle$, localized on sublattice $A$, is used to construct the projection gauge. 

\begin{figure}[t!]
\begin{center}
\includegraphics[width=3.4in]{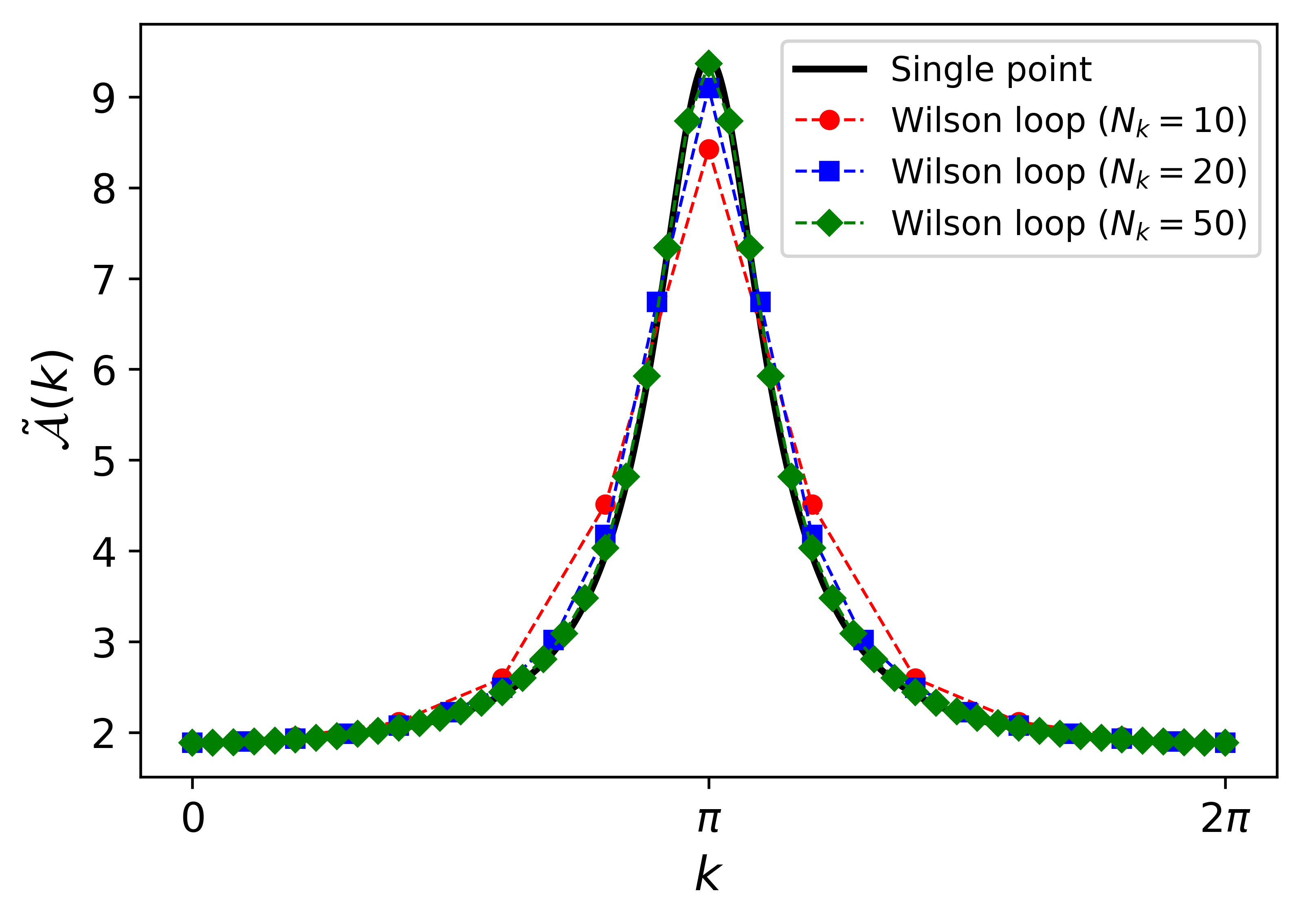}
\end{center}
\vspace{-7.5mm}
\caption{
Berry connection in the projection gauge as a function of crystal momentum (reduced coordinates) for $v=1$ and $w=1.5$. The solid black curve shows the single-point Berry connection. Dashed curves show the discrete connection obtained from Wilson loops at increasing mesh densities. As $N_k$ increases, the discrete connection converges to the single-point result.
}
\label{fig:3}
\end{figure}

Figure \ref{fig:3} shows the Berry connection in the projection gauge as a function of crystal momentum. The solid black curve corresponds to the single-point formulation developed here, which yields the exact Berry connection at each sampled $k$-point. The dashed curves show the discrete Berry connection extracted from overlap-based Wilson loops using \eq{discrete_connection}, evaluated in the same projection gauge. As the mesh density increases, the Wilson-loop connection converges uniformly to the single-point result. 

When the Berry phase is computed using overlap-based Wilson loops, convergence is rapid even on relatively coarse $k$-meshes. By contrast, evaluating the phase via numerical integration of the projection-gauge Berry connection exhibits slower convergence, particularly when the band gap is small. In this regime, the Berry connection develops a narrow peak in $k$, and a coarse mesh may not adequately resolve its structure, leading to quadrature error in the discrete integral. The convergence behavior of the two approaches is summarized in Table~\ref{tab:phases}.

This difference arises from the nature of the numerical integration rather than from any deficiency of the single-point construction. The Wilson-loop formulation discretizes parallel transport at the level of the holonomy, thereby capturing the accumulated geometric phase over each mesh segment. For the Berry phase, this is sufficient, since it depends only on the band-traced component of the connection and can therefore be expressed entirely in terms of the determinant of a product of overlap matrices.

The single-point approach instead provides the local Berry connection itself, whose integral must be approximated using standard quadrature. While this introduces mesh-dependent integration error, it provides direct access to the full connection matrix at each $k$-point. The advantage of local access to $\tilde{\mathcal{A}}_\mu$ becomes apparent in higher-dimensional settings, as in Sec.~\ref{sec:FKM}, where quantities of interest depend explicitly on the non-Abelian structure of the Berry connection, and where no simple gauge-covariant discretization is known that can reduce the problem to a product of overlaps.

\newcommand\Tstrut{\rule{0pt}{2.9ex}}  
\newcommand\Bstrut{\rule[-1.2ex]{0pt}{0pt}} 
\begin{table}[t!]
    \centering
    \begin{ruledtabular}
    \begin{tabular}{lcc}
    $N_k$& $\phi_{\mathrm{SP}}-\pi$ & $\phi_{\mathrm{FD}}-\pi$\\
     \hline \Tstrut
     10  & 5$\times 10^{-2}$ & 0 \\
     20  & 9$\times 10^{-4}$ & 0 \\
     50  & 5$\times 10^{-9}$ & 0 \\
     100 & 0 & 0 
    \end{tabular}
    \end{ruledtabular}
    \caption{Convergence of the Berry phase for the SSH model as a function of $k$-mesh density $N_k$. The table reports deviations from the quantized value $\pi$ obtained using the single-point (SP) projection-gauge formulation and the overlap-based finite-difference (FD) Wilson-loop approach. While the Wilson-loop method yields the exact quantized phase even on coarse meshes, the single-point method converges systematically as the mesh is refined.} 
    \label{tab:phases}
\end{table}

\section{Axion angle in the projection gauge}
\label{sec:FKM}

The Chern--Simons axion angle $\theta$ is a geometric property of the occupied Bloch states of a three-dimensional insulator and plays a central role in magnetoelectric response theory and in the topological classification of insulating phases. The axion angle can be expressed as a Brillouin-zone integral of the Chern--Simons 3-form constructed from the non-Abelian Berry connection. Explicitly, \cite{chern1974characteristic, Malashevich_2010, mong2010antiferromagnetic, essin2010orbital}
\begin{equation}
\label{eq:CS3form}
 \theta = -\frac{1}{4\pi} \int_{\text{BZ}} d^3k \,
\varepsilon^{\mu\nu\sigma} \mathrm{Tr} \left[
    \mathcal{A}_\mu \partial_\nu \mathcal{A}_\sigma
    - \frac{2i}{3} \mathcal{A}_\mu \mathcal{A}_\nu \mathcal{A}_\sigma
\right],
\end{equation}
where $\mathcal{A}_\mu$ is the non-Abelian Berry connection from \eq{berry_connection}. 

Because the Chern--Simons integrand is not gauge-covariant, the integral in \eqs{CS3form}{CS3form-curv} is invariant only modulo $2\pi$, giving $\theta$ the interpretation of an angular variable. In the presence of time-reversal ($\mathcal{T}$) or inversion ($\mathcal{P}$) symmetry, this ambiguity reduces to a quantization of $\theta$ in integer multiples of $\pi$, with odd (even) multiples corresponding to an odd (even) strong $\mathbb{Z}_2$ topological index in three dimensions \cite{essin2010orbital, qi2008topological, coh2011chern}. More generally, any cyclic adiabatic evolution of the Bloch Hamiltonian $H(\mathbf{k},\lambda)$ whose associated second Chern number in $(\mathbf{k},\lambda)$ space is nonzero will pump $\theta$ by an integer multiple of $2\pi$ \cite{essin2010orbital, taherinejad2015adiabatic}.

However, the direct numerical evaluation of $\theta$ is nontrivial \cite{coh2011chern, Liu2015, olsen2017surface, varnava2020axion}. The difficulty stems from the fact that---unlike the one-dimensional Berry phase---no fully gauge-invariant discretized expression for the Chern--Simons 3-form exists in terms of products of overlap matrices, except in the single-band case \cite{coh2011chern}. As a result, the construction of a smooth, periodic gauge and the direct evaluation of $\mathcal A_\mu$ become unavoidable prerequisites for computing $\theta$. The problem of constructing a smooth, periodic gauge is closely related to that of constructing well-localized Wannier functions. Accordingly, several approaches based on Wannier and hybrid Wannier representations have been developed to compute the Chern--Simons axion angle \cite{coh2011chern, olsen2017surface, varnava2020axion}. 

Here, we take a complementary approach. We combine the single-$k$ expression for the projection-gauge connection $\tilde{\mathcal{A}}_\mu(\mathbf{k}, \beta)$ derived in Sec.~\ref{sec:conn} with the Chern--Simons 3-form to compute the axion angle along an adiabatic cycle in the Fu--Kane--Mele (FKM) model. Once the projection gauge is fixed, $\tilde{\mathcal{A}}_\mu$ can be obtained directly at each $(\mathbf{k}, \beta)$ as a single-point quantity, without finite differences or overlap matrices between neighboring $k$ points. This eliminates branch-sensitive matrix logarithms and reduces sensitivity to gauge discontinuities, making it practical to evaluate the Chern--Simons integral directly. We will compare our method against the standard finite-difference approach and against a gauge-invariant formulation for \emph{changes} in $\theta$ along a cycle based on the 4-curvature.

\subsection{Model and adiabatic cycle}

The FKM model is a tight-binding model on a diamond lattice with spin-orbit coupling that realizes a three-dimensional topological insulator with time-reversal ($\mathcal{T}$) and inversion ($\mathcal{P}$) symmetries. Each of these symmetries, in combination with the model's nontrivial topological nature, quantizes the axion angle to $\theta=\pi$. 

To break the symmetries that quantize $\theta$, we follow Ref.~\cite{essin2010orbital} by introducing an adiabatic parameter $\beta$ that enters the Hamiltonian through both a staggered Zeeman field and a modulation of the nearest-neighbor hopping amplitudes along the $[111]$ direction. The model Hamiltonian, written in second-quantized form, is 
\begin{align}
\label{eq:FKM_H}
H(\beta) =  \; &  \sum_{\langle ij \rangle} t_{ij}(\beta) c_i^{\dagger} c_j
+ \mathbf{h}(\beta) \cdot \left( \sum_{i \in \text{A}} c_i^\dagger \mathbf{\sigma} c_i  - \sum_{i \in \text{B}} c_i^\dagger \mathbf{\sigma} c_i \right) +\\ & \nonumber i\lambda_{\text{SO}}\sum_{\langle\langle ij \rangle\rangle} c_i^{\dagger} \boldsymbol{\sigma} \cdot (\mathbf{d}_{ij}^{1} \times \mathbf{d}_{ij}^{2}) c_j  ,
\end{align}
where $\mathbf{d}_{ij}^{1,2}$ are the two nearest-neighbor bond vectors connecting sites $i$ and $j$, $\boldsymbol{\sigma}$ denotes the vector of Pauli matrices acting on spin, and $A$ and $B$ label the two sublattices. The staggered Zeeman field breaks time-reversal symmetry and is chosen as $\mathbf{h}(\beta)= m\sin\beta\, \hat{n}$, with $\hat{n}$ pointing along the $[111]$ direction. The hopping modulation breaks inversion symmetry and is taken to be $t_{ij} = 3t+m\cos\beta$ for bonds oriented along $[111]$, and $t_{ij}=t$ otherwise. With this parameterization, the bulk energy gap remains open throughout the adiabatic cycle as $\beta$ is varied from $0$ to $2\pi$. Unless otherwise specified, all $k$-meshes are uniform and of size $ 50 \times 50 \times 50$.

\subsection{Gauge-invariant 4-curvature}
\label{sec:4curv}
As an independent benchmark of our single-point formulation, we compute the axion angle using the gauge-invariant four-curvature expression, which determines the \emph{change} in $\theta$ along an adiabatic cycle. Treating the cyclic parameter $\beta$ as an additional coordinate, the occupied bands define a vector bundle over the four-torus $(k_x,k_y,k_z,\beta)$. The axion response can be written as \cite{qi2008topological}
\begin{equation}
\label{eq:4curv}
    \theta(\beta) -\theta(0) = \frac{1}{16 \pi} \int_0^{\beta} d\beta^\prime \int_{\text{BZ}} d^3k\, \varepsilon^{\mu\nu\rho\sigma} \text{Tr} \left[\Omega_{\mu\nu} \Omega_{\rho\sigma} \right],
\end{equation}
where $\Omega_{\mu\nu}$ is the non-Abelian Berry curvature of the occupied subspace and $(\mu,\nu,\rho,\sigma) \in \{k_x,k_y,k_z,\beta\}$. The integral over the 4-torus manifold is quantized and equals the second Chern number $C_2$ of the bundle. For the adiabatic cycle considered here, the associated second Chern number is $C_2 = 1$, implying that the axion angle is pumped by $2\pi$ over one full cycle \cite{essin2009magnetoelectric,taherinejad2015adiabatic, essin2010orbital},
\begin{equation}
    \theta(2\pi) - \theta(0) = 2\pi\,.
\end{equation}

Since Eq.~\eqref{eq:4curv} depends only on the occupied projector defined over the four-dimensional parameter space, it is manifestly gauge invariant and requires no choice of smooth Bloch-state gauge. At $\beta=0$, inversion and time-reversal symmetry quantize the axion angle to $0$ or $\pi$, and since the system is topologically trivial at this parameter value, $\theta(0)=0$. In this case, \eq{4curv} therefore gives us $\theta(\beta)$ directly, and its computed values are shown as the dashed black curve in Fig.~\ref{fig:1}.

\subsection{Chern--Simons 3-form in projection gauge}

We now compute the axion angle at fixed $\beta$ using the Chern--Simons 3-form, Eq.~\ref{eq:CS3form-curv}, evaluated with the projection-gauge non-Abelian Berry connection $\tilde{\mathcal{A}}_\mu$ derived analytically in Sec.~\ref{sec:conn}, and compare it to the standard finite difference approach. This provides a direct test of both the numerical stability and internal consistency of our fixed-$k$ formulation.

\begin{figure}[t!]
\begin{center}
\includegraphics[width=3.4in]{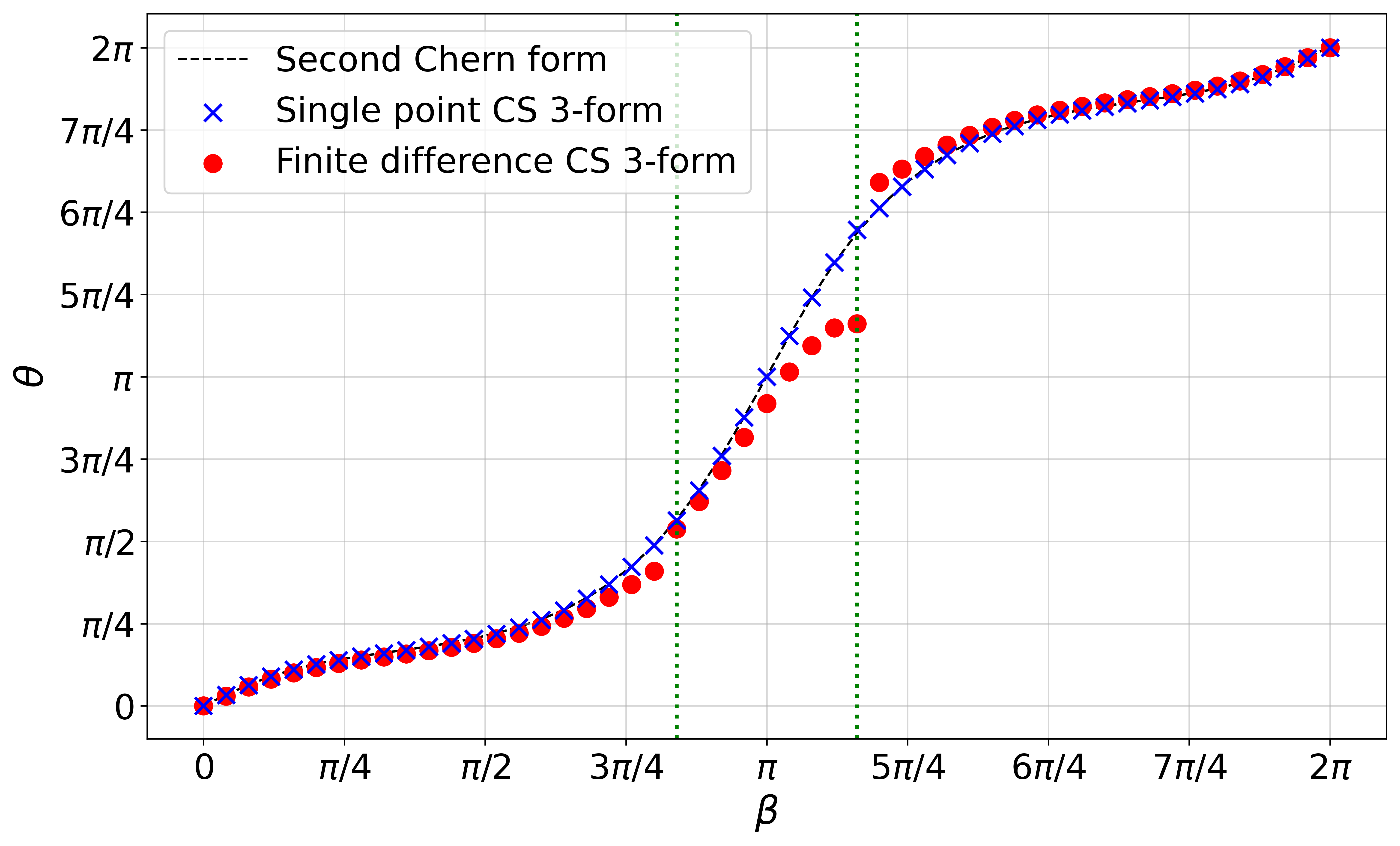}
\end{center}
\vspace{-7.5mm}
\caption{The axion angle $\theta(\beta)$ over one adiabatic cycle $\beta \in [0, 2\pi]$. We compute $\theta$ using three independent approaches: (i) the gauge-invariant formulation based on integrating the second Chern form (four-curvature),  shown as a dotted black line; (ii) the Chern--Simons 3-form from \eq{CS3form-curv} using the analytic single-point Berry connection of \eq{tildeA_final}, plotted in blue; and (iii) the same Chern--Simons expression evaluated using the finite difference connection of \eq{fd_connection}, shown in red. The vertical green dotted lines mark the region within which we use trial wavefunctions that do not respect $\mathcal{T}$. The nontrivial second Chern number ($C_2 = 1$) produces a net $2\pi$ shift of $\theta$ over one adiabatic cycle.}
\label{fig:1}
\end{figure}

We build the projection gauge from two trial wavefunctions projected onto the occupied bands. In the topologically trivial phase, the trial wavefunctions take the form of a Kramers pair, with each being spin polarized and having equal weight on the tight-binding orbitals situated on the $A$ and $B$ sublattices,
\begin{align}
    \label{eq:trial_triv}
    |t_1\rangle &= \frac{1}{\sqrt{2}}\left(|A\uparrow\rangle - |B\uparrow\rangle \right)\\
    |t_2 \rangle &= \frac{1}{\sqrt{2}}\left(|A\downarrow\rangle - |B\downarrow\rangle \right) .
\end{align}
At $\beta=\pi$, the system is in a topologically nontrivial phase, and a smooth, periodic gauge respecting time-reversal symmetry does not exist. To avoid the associated obstruction \cite{soluyanov2011}, we therefore break $\mathcal{T}$ in the gauge by employing trial wavefunctions whose spin structures are no longer related to each other by $\mathcal{T}$,
\begin{align}
    \label{eq:trial_nontriv}
    |t_1\rangle &= \frac{1}{\sqrt{2}}\left(|A\uparrow\rangle - |A \downarrow\rangle \right),\\
    \label{eq:trial_nontriv2}
    |t_2 \rangle &= \frac{1}{\sqrt{2}}\left(|B\uparrow\rangle - |B\downarrow\rangle \right).
\end{align}
These non-Kramers trial states are used only within a narrow window around $\beta=\pi$, indicated by the green dotted lines in Fig.~\ref{fig:1}, to avoid numerical instabilities associated with small singular values of the overlap matrix.

To avoid computing $\partial_\nu \mathcal{A}_\sigma$ in \eq{CS3form} by finite differences, we eliminate it in favor of the gauge-covariant non-Abelian Berry curvature
\begin{equation}
    \Omega_{\mu\nu, mn} = \partial_{\mu} \mathcal{A}_{\nu, mn} - \partial_{\nu} \mathcal{A}_{\mu,mn} - i[\mathcal{A}_\mu, \mathcal{A}_{\nu}]_{mn}
   \end{equation}
so that
\begin{equation}
\label{eq:CS3form-curv}
    \theta = -\frac{1}{4\pi} \int_{\text{BZ}} d^3 k\, \varepsilon^{\mu\nu\sigma} \text{Tr}\left[ \frac{1}{2}\mathcal{A}_{\mu}\Omega_{\nu\sigma} + \frac{i}{3} \mathcal{A}_\mu \mathcal{A}_\nu \mathcal{A}_\sigma\right].
\end{equation}
The advantage of this reformulation is that $\Omega$ can be computed using the Kubo formula
\begin{align}
\label{eq:bc-def}
    \Omega_{\mu\nu, mn} &=\sum_{c}^{\mathrm{empty}} \frac{\langle u_{m\mathbf{k}}|\partial_{\mu} H_{\mathbf{k}}|u_{c\mathbf{k}}\rangle \langle u_{c\mathbf{k}}|\partial_{\nu} H_{\mathbf{k}}|u_{n\mathbf{k}}\rangle}{(E_{m\mathbf{k}}-E_{c\mathbf{k}})(E_{n\mathbf{k}} -E_{c\mathbf{k}})} - (\mu \leftrightarrow \nu) \\ &= (\mathcal{D}^H_\mu \mathcal{D}_\nu^{H\dagger} - \mathcal{D}^H_\nu \mathcal{D}_\mu^{H\dagger})_{mn}.
\end{align}
In the first-principles context, this is approximated by a sum over all conduction bands included in the calculation, whereas in tight-binding it is approximated by a simple sum over a finite number of empty bands. 

From the trial states, we construct the overlap matrices defined in \eqs{ovlp}{Sperp} and follow the procedure of Sec.~\ref{sec:conn} to obtain the projection-gauge connection $\tilde{\mathcal{A}}_\mu$. Notably, the explicit rotation into the projection gauge is never carried out; all quantities are computed directly from the overlap matrices. The resulting connection, \eq{tildeA_final}, is inserted into the Chern--Simons 3-form~\eqref{eq:CS3form-curv} to yield $\theta(\beta)$, shown as the blue curve in Fig.~\ref{fig:1}.

The agreement with the gauge-invariant four-curvature result is excellent over the entire cycle. At $\beta=\pi$, symmetry enforces $\theta=\pi$, and our numerical value deviates from this by less than $10^{-6}$, with the residual error attributable to the Brillouin-zone discretization of the integral and finite numerical precision.

\subsection{Comparison with finite-difference connections}

For comparison, we also compute the axion angle using the standard finite-difference construction of the non-Abelian Berry connection. The procedure is as follows:
(i) we construct the projected and orthonormalized Bloch-like states $\ket{\tilde{u}_{n\mathbf{k}}}$ from \eq{tildepsi_from_phi},
(ii) we form overlap matrices between neighboring $k$-points,
\begin{equation}
    M_\mu(\mathbf{k})_{mn} = \langle \tilde{u}_{m\mathbf{k}}|\tilde{u}_{n,\mathbf{k } + \delta k \hat{\mu}}\rangle
\end{equation}
where $\hat{\mu}$ is a Cartesian unit vector and $\delta k$ is the mesh spacing, (iii) we extract the unitary part of the overlap matrix via the polar decomposition\footnote{In practice, the polar decomposition is obtained from the singular value decomposition $M=V\Sigma W^\dagger$, with $\mathcal U = V W^\dagger$ and $\mathcal{P} =(M^\dagger M)^{1/2} = W\Sigma W^\dagger$.},
\begin{equation}
        M_{\mu} = \mathcal{U}_{\mu}\mathcal{P}_\mu
\end{equation}
where $\mathcal U_\mu$ is unitary and $\mathcal P_\mu$ is positive definite,
and (iv) we evaluate the discretized non-Abelian Berry connection matrix as
\begin{equation}
\label{eq:fd_connection}
    \tilde{\mathcal{A}}_\mu = \frac{i}{\delta k} \ln\left[\,\mathcal{U}_\mu \right]
\end{equation}
where the matrix logarithm is taken with a continuous branch choice. For sufficiently small $\delta k$, this approximates the continuum connection of \eq{berry_connection} to $\mathcal{O}(\delta k)$. This connection is then inserted into the Chern--Simons 3-form of \eq{CS3form-curv} to evaluate $\theta(\beta)$. The resulting axion angle is shown as the red curve in Fig.~\ref{fig:1}. 

Near the topological obstruction at $\beta=\pi$ (Fig.~\ref{fig:1}), the finite-difference construction becomes numerically fragile. The overlap matrices develop small singular values, making the polar decomposition ill-conditioned and causing the logarithm of the unitary link matrices to fluctuate on coarse meshes, even when $\mathcal{T}$-breaking trial states are used. The rapid variation of the projection-gauge states $|\tilde{u}_{nk}\rangle$ in this region is the underlying cause: as the Bloch-like states change substantially between neighboring $k$-points, the discretized connection fails to resolve the true geometric structure.

\begin{figure}[t!]
\begin{center}
\includegraphics[width=3.4in]{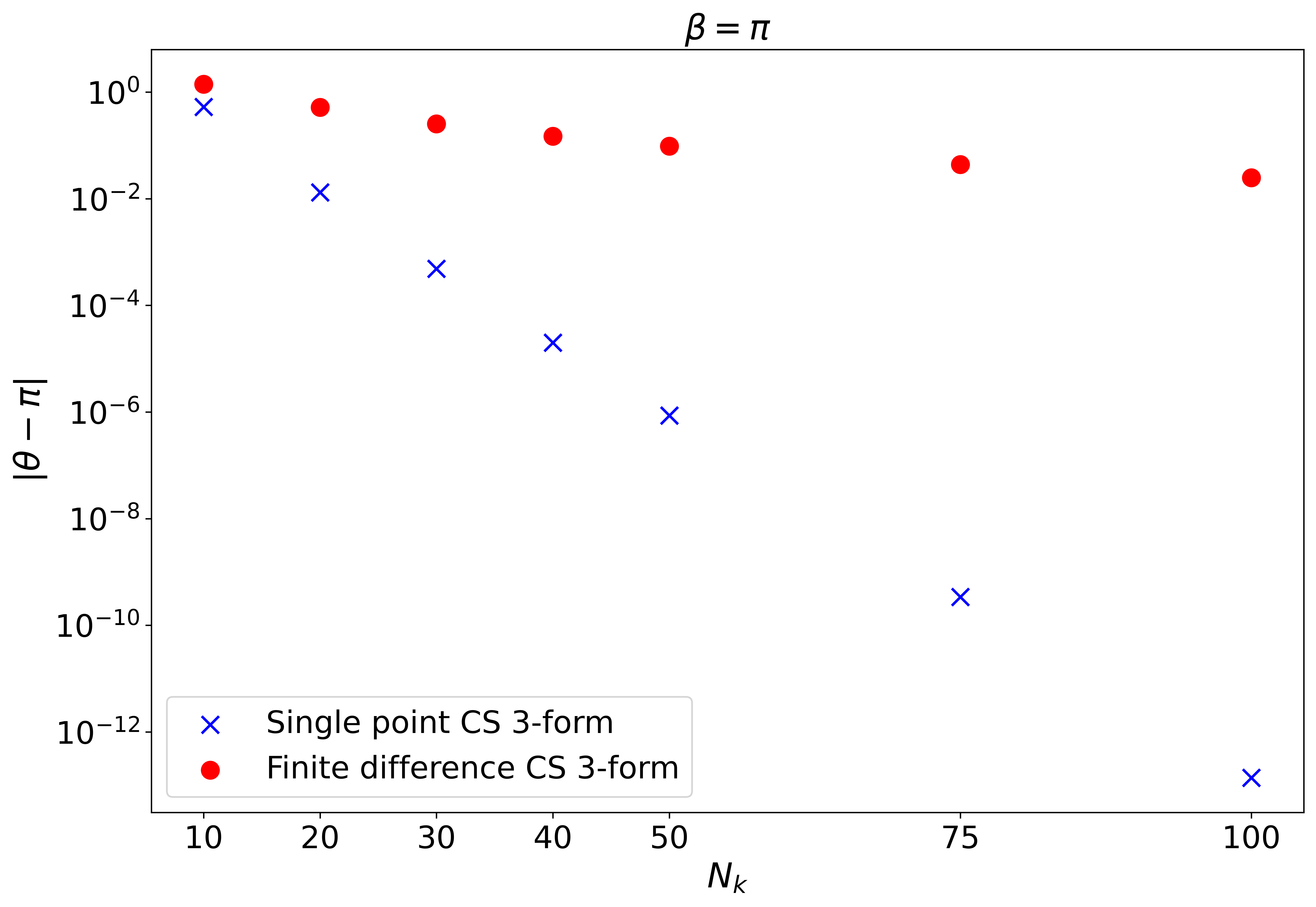}
\end{center}
\vspace{-7.5mm}
\caption{
The convergence of $\theta$ toward its quantized value of $\pi$, when $\beta=\pi$, with respect to the $k$-mesh density. The horizontal axis denotes the number of $k$-points along each reciprocal lattice vector; e.g., a given $N_{k}$ corresponds to an $N_k \times N_k \times N_k$-sized $k$-mesh. The single-point formulation rapidly approaches the continuum result, whereas the finite-difference construction converges more slowly due to additional discretization error in the Berry connection itself. 
}
\label{fig:2}
\end{figure}

The resulting discrepancy between the finite-difference result (red) and the single-point result (blue) in Fig.~\ref{fig:1} is therefore one of underconvergence at finite mesh density, not of a fundamental limitation of the finite-difference approach. By contrast, the single-point formulation provides an exact expression for the non-Abelian Berry connection at each $k$ point, so the only approximation arises from the first-order quadrature error in the Brillouin-zone integration. As a result, convergence is rapid once the $k$ mesh resolves the relevant geometric structure, as shown in Fig~\ref{fig:2}.

Figure 3 highlights the consequences for convergence. Although both approaches share the same first-order quadrature error in the Brillouin-zone integration, the additional discretization at the level of the connection leads to substantially slower convergence for the finite-difference method. This behavior reflects a structural feature of the method: the Berry connection itself is approximated through neighboring-$k$ overlaps, so discretization error is introduced before the Chern--Simons integral is evaluated.

\section{Discussion}

We have derived an exact Kubo-like expression for the non-Abelian Berry connection in the projection (Wannier) gauge. By replacing finite-difference derivatives with a strictly $k$-local formulation, the method removes errors associated with the discretized momentum derivatives and branch-sensitive matrix logarithms. In one dimension, the SSH model provides a controlled benchmark: in the fine-mesh limit, the derived $\tilde{\mathcal{A}}$ reproduces the quantized Berry phase and agrees with the standard parallel-transport construction. On coarse meshes, parallel transport remains more robust for evaluating the Berry phase, reflecting the special role of gauge-invariant holonomy in one dimension. 

In higher dimensions, however, no analogous gauge-invariant discretization exists for geometric quantities that depend explicitly on the Berry connection itself. In this regime, construction of $\tilde{\mathcal{A}}_\mu$ in a smooth gauge is unavoidable, and the present local formulation provides a clear advantage. We demonstrated this in three dimensions by evaluating the Chern--Simons 3-form in the FKM model. Comparison with finite-difference and gauge-invariant 4-curvature approaches shows that the single-$k$ formulation yields the correct axion angle over an adiabatic cycle and exhibits improved convergence with increasing mesh density. 
 
As noted in Sec.~\ref{sec:conn}, in a tight-binding setting Eq.~\ref{eq:tildeA_final} yields the Berry connection exactly, while in a first-principles setting the Sternheimer equation provides a route to evaluating the interband sum without explicit empty states at moderate additional cost. This contrasts with the finite-difference overlap scheme, which introduces an additional discretization error at the level of the connection itself. Whether the single-point construction is preferable depends on the quantity being computed: for quantities that require the full non-Abelian $\tilde{\mathcal{A}}_\mu$ in a smooth gauge and for which finite-difference errors are not marginal, the single-point formulation offers a favorable trade-off. The Chern--Simons 3-form entering the axion angle $\theta$ is the primary example of such a quantity: its integrand involves $\mathcal{A}_\mu$ and has no known reformulation in terms of Wilson loops or gauge-covariant quantities alone, and has proven difficult to control in practice \cite{coh2011chern, Liu2015, olsen2017surface, varnava2020axion}. Most other geometric and response quantities---including the Berry curvature, quantum metric tensor, optical conductivity, and shift current \cite{ponce2021, azpiroz2018}---admit Kubo-like or gauge-covariant expressions that do not require a globally smooth gauge. Even for response quantities that do not require a smooth gauge, the off-diagonal position matrix elements in the Wannier basis are conventionally obtained from finite-difference overlaps. The standard interpolation scheme can introduce non-negligible errors in these matrix elements, motivating the development of self-consistent \cite{thummler2026} and higher-order equivariant \cite{lihm2026} alternatives. The exact single-point construction provides a complementary approach that avoids the finite-difference step entirely and may improve the accuracy of these underlying matrix elements.

Although not yet implemented in a fully self-consistent first-principles workflow, the derivation is completely general and directly applicable in that setting. Because the projection-gauge Berry connection is obtained as a strictly local object in momentum space, the method avoids reliance on neighboring-$k$ overlaps and large overlap matrices. This locality makes the approach naturally compatible with symmetry-reduced meshes, nonuniform or adaptive grids, and parallel implementations in which each $k$-point is treated independently. While adequate $k$-point sampling remains essential for Brillouin-zone integrals, the present approach removes a distinct source of numerical instability inherent to finite-difference constructions. Moreover, because Wannier interpolation relies on the off-diagonal position matrix elements in the Wannier basis, the availability of a closed-form expression for the projection-gauge Berry connection may enable alternative interpolation strategies in which geometric quantities are constructed directly at each $k$ from local information, rather than through post-processed overlap matrices. Incorporating the present framework into a fully first-principles Wannier-based implementation is therefore a natural and promising direction for future work.

\vspace{0.6cm}

\acknowledgments
This work was supported by NSF Grant DMR-2421895. We thank Andrea Urru and Daniel Seleznev for helpful discussions related to this work.

\appendix

\section{Deriving the Sylvester equation solution}
\label{apdx:sylvester}
We have a Sylvester equation for $\partial_\mu \mathcal{O}^{1/2}$ of the form,
\begin{equation}
    \label{eq:slyv_eq_apdx}
    \mathcal{O}^{1/2}(\partial_\mu \mathcal{O}^{1/2})+(\partial_\mu \mathcal{O}^{1/2})\mathcal{O}^{1/2}=\partial_\mu \mathcal{O}.
\end{equation}
We can solve this by transforming into the eigenbasis of $\mathcal{O}^{1/2}$. Since $\mathcal{O}^{1/2}$ is Hermitian by definition, this transformation always exists, $\mathcal{O}^{1/2}= V \Sigma V^{\dagger}$ where $V$ is the unitary of eigenvectors and $\Sigma = \textrm{diag}(\sigma_1, \sigma_2, \cdots )$ is the matrix of eigenvalues. Substituting the spectral decomposition into \eq{slyv_eq_apdx} we get, 
\begin{equation}
\label{eq:eigdecomp}
    (V\Sigma V^{\dagger})(\partial_\mu \mathcal{O}^{1/2}) + (\partial_\mu \mathcal{O}^{1/2}) (V \Sigma V^{\dagger} ) = \partial_\mu \mathcal{O}.
\end{equation}
We multiply $V^{\dagger}$ from the left and $V$ from the right, obtaining,
\begin{align}
    \label{eq:sylv_step_2}
    &\Sigma \,(V^{\dagger}\,\partial_\mu \mathcal{O}^{1/2}\,V)
    + (V^{\dagger}\,\partial_\mu \mathcal{O}^{1/2}\,V)\,\Sigma = 
   V^{\dagger} (\partial_\mu \mathcal{O})V.
\end{align}
Since $\Sigma_{mn} = \sigma_m\delta_{mn}$, the component form is
\begin{equation}
\label{eq:Pprime_comp}
    (\sigma_m+\sigma_n)
    (V^{\dagger}\,\partial_\mu \mathcal{O}^{1/2}\,V)_{mn}
    = \left( V^{\dagger} (\partial_\mu \mathcal{O})V \right)_{mn}.
\end{equation}
We divide by the singular values, followed by a multiplication of $V$ from the left and $V^{\dagger}$ on the right to get,
\begin{equation}
\label{eq:sylv_sol}
    (\partial_\mu \mathcal{O}^{1/2})_{mn}
    =
    V_{mi}
    \left[ 
    \frac{
        \left(
           V^{\dagger} (\partial_\mu \mathcal{O})V
        \right)_{ij}
    }{
        \sigma_i+\sigma_j
    }\right] V^{\dagger}_{jn}.
\end{equation}
Here, the division by $(\sigma_i+\sigma_j)$ is understood elementwise in the $\Sigma$–diagonal basis.

\bibliography{ref}

@software{cole_2026_18727385,
  author       = {Cole, Trey and
                  Vanderbilt, David},
  title        = {Exact expression for the Berry connection in the
                   projection gauge (Zenodo)
                  },
  month        = feb,
  year         = 2026,
  publisher    = {Zenodo},
  doi          = {10.5281/zenodo.18727385},
  url          = {https://doi.org/10.5281/zenodo.18727385},
   note         = {available at https://doi.org/10.5281/zenodo.18727385},
  swhid        = {swh:1:dir:d496944f2ca5786dead5c9e470872596a2098012
                   ;origin=https://doi.org/10.5281/zenodo.18727384;vi
                   sit=swh:1:snp:a7b958ccc01d83484ac9f3b19bd674984433
                   6095;anchor=swh:1:rel:cb2e7b67ea68a8a7d7df59d30273
                   32e5a61f3619;path=projection-connection-1.0-arxiv
                  },
}

@misc{thummler2026,
      title={Self-consistent evaluation of the Berry connection for Wannier functions}, 
      author={Martin Thümmler and Alexander Croy and Thomas Lettau and Ulf Peschel and Stefanie Gräfe},
      year={2026},
      eprint={2604.21660},
      archivePrefix={arXiv},
      primaryClass={cond-mat.mtrl-sci},
      url={https://arxiv.org/abs/2604.21660}, 
}

@misc{lihm2026,
      title={Accurate calculation of Wannier centers, position matrix, and composite operators using translationally equivariant and higher-order finite differences}, 
      author={Jae-Mo Lihm and Minsu Ghim and Seung-Ju Hong and Cheol-Hwan Park},
      year={2026},
      eprint={2604.22614},
      archivePrefix={arXiv},
      primaryClass={cond-mat.mtrl-sci},
      url={https://arxiv.org/abs/2604.22614}, 
}

@article{Baroni1987,
  title = {Green's-function approach to linear response in solids},
  author = {Baroni, Stefano and Giannozzi, Paolo and Testa, Andrea},
  journal = {Phys. Rev. Lett.},
  volume = {58},
  issue = {18},
  pages = {1861--1864},
  numpages = {0},
  year = {1987},
  month = {May},
  publisher = {American Physical Society},
  doi = {10.1103/PhysRevLett.58.1861},
  url = {https://link.aps.org/doi/10.1103/PhysRevLett.58.1861}
}

@article{Gonze1995,
  title = {Adiabatic density-functional perturbation theory},
  author = {Gonze, Xavier},
  journal = {Phys. Rev. A},
  volume = {52},
  issue = {2},
  pages = {1096--1114},
  numpages = {0},
  year = {1995},
  month = {Aug},
  publisher = {American Physical Society},
  doi = {10.1103/PhysRevA.52.1096},
  url = {https://link.aps.org/doi/10.1103/PhysRevA.52.1096}
}

@article{Gonze1997,
  title = {Dynamical matrices, Born effective charges, dielectric permittivity tensors, and interatomic force constants from density-functional perturbation theory},
  author = {Gonze, Xavier and Lee, Changyol},
  journal = {Phys. Rev. B},
  volume = {55},
  issue = {16},
  pages = {10355--10368},
  numpages = {0},
  year = {1997},
  month = {Apr},
  publisher = {American Physical Society},
  doi = {10.1103/PhysRevB.55.10355},
  url = {https://link.aps.org/doi/10.1103/PhysRevB.55.10355}
}

@article{baroni2001,
  title = {Phonons and related crystal properties from density-functional perturbation theory},
  author = {Baroni, Stefano and de Gironcoli, Stefano and Dal Corso, Andrea and Giannozzi, Paolo},
  journal = {Rev. Mod. Phys.},
  volume = {73},
  issue = {2},
  pages = {515--562},
  numpages = {0},
  year = {2001},
  month = {Jul},
  publisher = {American Physical Society},
  doi = {10.1103/RevModPhys.73.515},
  url = {https://link.aps.org/doi/10.1103/RevModPhys.73.515}
}

@article{Liu2015,
  title = {Gauge-discontinuity contributions to Chern-Simons orbital magnetoelectric coupling},
  author = {Liu, Jianpeng and Vanderbilt, David},
  journal = {Phys. Rev. B},
  volume = {92},
  issue = {24},
  pages = {245138},
  numpages = {15},
  year = {2015},
  month = {Dec},
  publisher = {American Physical Society},
  doi = {10.1103/PhysRevB.92.245138},
  url = {https://link.aps.org/doi/10.1103/PhysRevB.92.245138}
}

@article{ponce2021,
  title = {First-principles predictions of Hall and drift mobilities in semiconductors},
  author = {Ponc\'e, Samuel and Macheda, Francesco and Margine, Elena Roxana and Marzari, Nicola and Bonini, Nicola and Giustino, Feliciano},
  journal = {Phys. Rev. Res.},
  volume = {3},
  issue = {4},
  pages = {043022},
  numpages = {36},
  year = {2021},
  month = {Oct},
  publisher = {American Physical Society},
  doi = {10.1103/PhysRevResearch.3.043022},
  url = {https://link.aps.org/doi/10.1103/PhysRevResearch.3.043022}
}

@article{azpiroz2018,
  title = {Ab initio calculation of the shift photocurrent by Wannier interpolation},
  author = {Iba\~nez-Azpiroz, Julen and Tsirkin, Stepan S. and Souza, Ivo},
  journal = {Phys. Rev. B},
  volume = {97},
  issue = {24},
  pages = {245143},
  numpages = {13},
  year = {2018},
  month = {Jun},
  publisher = {American Physical Society},
  doi = {10.1103/PhysRevB.97.245143},
  url = {https://link.aps.org/doi/10.1103/PhysRevB.97.245143}
}

@article{wang2006,
  title = {Ab initio calculation of the anomalous Hall conductivity by Wannier interpolation},
  author = {Wang, Xinjie and Yates, Jonathan R. and Souza, Ivo and Vanderbilt, David},
  journal = {Phys. Rev. B},
  volume = {74},
  issue = {19},
  pages = {195118},
  numpages = {15},
  year = {2006},
  month = {Nov},
  publisher = {American Physical Society},
  doi = {10.1103/PhysRevB.74.195118},
  url = {https://link.aps.org/doi/10.1103/PhysRevB.74.195118}
}

@article{yates2007,
  title = {Spectral and Fermi surface properties from Wannier interpolation},
  author = {Yates, Jonathan R. and Wang, Xinjie and Vanderbilt, David and Souza, Ivo},
  journal = {Phys. Rev. B},
  volume = {75},
  issue = {19},
  pages = {195121},
  numpages = {11},
  year = {2007},
  month = {May},
  publisher = {American Physical Society},
  doi = {10.1103/PhysRevB.75.195121},
  url = {https://link.aps.org/doi/10.1103/PhysRevB.75.195121}
}

@article{MOSTOFI20142309,
title = {An updated version of wannier90: A tool for obtaining maximally-localised Wannier functions},
journal = {Computer Physics Communications},
volume = {185},
number = {8},
pages = {2309-2310},
year = {2014},
issn = {0010-4655},
doi = {https://doi.org/10.1016/j.cpc.2014.05.003},
url = {https://www.sciencedirect.com/science/article/pii/S001046551400157X},
author = {Arash A. Mostofi and Jonathan R. Yates and Giovanni Pizzi and Young-Su Lee and Ivo Souza and David Vanderbilt and Nicola Marzari},
}

@article{resta1994,
  title = {Macroscopic polarization in crystalline dielectrics: the geometric phase approach},
  author = {Resta, Raffaele},
  journal = {Rev. Mod. Phys.},
  volume = {66},
  issue = {3},
  pages = {899--915},
  numpages = {0},
  year = {1994},
  month = {Jul},
  publisher = {American Physical Society},
  doi = {10.1103/RevModPhys.66.899},
  url = {https://link.aps.org/doi/10.1103/RevModPhys.66.899}
}

@article{vanderbilt1993,
  title = {Electric polarization as a bulk quantity and its relation to surface charge},
  author = {Vanderbilt, David and King-Smith, R. D.},
  journal = {Phys. Rev. B},
  volume = {48},
  issue = {7},
  pages = {4442--4455},
  numpages = {0},
  year = {1993},
  month = {Aug},
  publisher = {American Physical Society},
  doi = {10.1103/PhysRevB.48.4442},
  url = {https://link.aps.org/doi/10.1103/PhysRevB.48.4442}
}

@article{marazzo2024,
  title = {Wannier-function software ecosystem for materials simulations},
  author = {Marrazzo, Antimo and Beck, Sophie and Margine, Elena R. and Marzari, Nicola and Mostofi, Arash A. and Qiao, Junfeng and Souza, Ivo and Tsirkin, Stepan S. and Yates, Jonathan R. and Pizzi, Giovanni},
  journal = {Rev. Mod. Phys.},
  volume = {96},
  issue = {4},
  pages = {045008},
  numpages = {54},
  year = {2024},
  month = {Dec},
  publisher = {American Physical Society},
  doi = {10.1103/RevModPhys.96.045008},
  url = {https://link.aps.org/doi/10.1103/RevModPhys.96.045008}
}

@book{Vanderbilt_2018, place={Cambridge}, title={Berry Phases in Electronic Structure Theory: Electric Polarization, Orbital Magnetization and Topological Insulators}, publisher={Cambridge University Press}, author={Vanderbilt, David}, year={2018}}

@article{xiao2010,
  title = {Berry phase effects on electronic properties},
  author = {Xiao, Di and Chang, Ming-Che and Niu, Qian},
  journal = {Rev. Mod. Phys.},
  volume = {82},
  issue = {3},
  pages = {1959--2007},
  numpages = {0},
  year = {2010},
  month = {Jul},
  publisher = {American Physical Society},
  doi = {10.1103/RevModPhys.82.1959},
  url = {https://link.aps.org/doi/10.1103/RevModPhys.82.1959}
}

@article{souza2001,
  title = {Maximally localized Wannier functions for entangled energy bands},
  author = {Souza, Ivo and Marzari, Nicola and Vanderbilt, David},
  journal = {Phys. Rev. B},
  volume = {65},
  issue = {3},
  pages = {035109},
  numpages = {13},
  year = {2001},
  month = {Dec},
  publisher = {American Physical Society},
  doi = {10.1103/PhysRevB.65.035109},
  url = {https://link.aps.org/doi/10.1103/PhysRevB.65.035109}
}

@article{qi2008topological,
  title={Topological field theory of time-reversal invariant insulators},
  author={Qi, Xiao-Liang and Hughes, Taylor L and Zhang, Shou-Cheng},
  journal={Physical Review B—Condensed Matter and Materials Physics},
  volume={78},
  number={19},
  pages={195424},
  year={2008},
  publisher={APS}
}

@article{essin2009magnetoelectric,
  title={Magnetoelectric polarizability and axion electrodynamics in crystalline insulators},
  author={Essin, Andrew M and Moore, Joel E and Vanderbilt, David},
  journal={Physical review letters},
  volume={102},
  number={14},
  pages={146805},
  year={2009},
  publisher={APS}
}

@article{chern1974characteristic,
  title={Characteristic forms and geometric invariants},
  author={Chern, Shiing-Shen and Simons, James},
  journal={Annals of Mathematics},
  volume={99},
  number={1},
  pages={48--69},
  year={1974},
  publisher={JSTOR}
}

@article{marzari1997,
  title = {Maximally localized generalized Wannier functions for composite energy bands},
  author = {Marzari, Nicola and Vanderbilt, David},
  journal = {Phys. Rev. B},
  volume = {56},
  issue = {20},
  pages = {12847--12865},
  numpages = {0},
  year = {1997},
  month = {Nov},
  publisher = {American Physical Society},
  doi = {10.1103/PhysRevB.56.12847},
  url = {https://link.aps.org/doi/10.1103/PhysRevB.56.12847}
}

@article{Malashevich_2010,
doi = {10.1088/1367-2630/12/5/053032},
url = {https://dx.doi.org/10.1088/1367-2630/12/5/053032},
year = {2010},
month = {may},
publisher = {},
volume = {12},
number = {5},
pages = {053032},
author = {Malashevich, Andrei and Souza, Ivo and Coh, Sinisa and Vanderbilt, David},
title = {Theory of orbital magnetoelectric response},
journal = {New Journal of Physics}
}

@article{essin2010orbital,
  title={Orbital magnetoelectric coupling in band insulators},
  author={Essin, Andrew M and Turner, Ari M and Moore, Joel E and Vanderbilt, David},
  journal={Physical Review B—Condensed Matter and Materials Physics},
  volume={81},
  number={20},
  pages={205104},
  year={2010},
  publisher={APS}
}

@article{mong2010antiferromagnetic,
  title={Antiferromagnetic topological insulators},
  author={Mong, Roger SK and Essin, Andrew M and Moore, Joel E},
  journal={Physical Review B—Condensed Matter and Materials Physics},
  volume={81},
  number={24},
  pages={245209},
  year={2010},
  publisher={APS}
}

@article{coh2011chern,
  title={Chern-Simons orbital magnetoelectric coupling in generic insulators},
  author={Coh, Sinisa and Vanderbilt, David and Malashevich, Andrei and Souza, Ivo},
  journal={Physical Review B—Condensed Matter and Materials Physics},
  volume={83},
  number={8},
  pages={085108},
  year={2011},
  publisher={APS}
}

@article{taherinejad2015adiabatic,
  title={Adiabatic pumping of Chern-Simons axion coupling},
  author={Taherinejad, Maryam and Vanderbilt, David},
  journal={Physical review letters},
  volume={114},
  number={9},
  pages={096401},
  year={2015},
  publisher={APS}
}

@article{olsen2017surface,
  title={Surface theorem for the Chern-Simons axion coupling},
  author={Olsen, Thomas and Taherinejad, Maryam and Vanderbilt, David and Souza, Ivo},
  journal={Physical Review B},
  volume={95},
  number={7},
  pages={075137},
  year={2017},
  publisher={APS}
}

@article{varnava2020axion,
  title={Axion coupling in the hybrid Wannier representation},
  author={Varnava, Nicodemos and Souza, Ivo and Vanderbilt, David},
  journal={Physical Review B},
  volume={101},
  number={15},
  pages={155130},
  year={2020},
  publisher={APS}
}

@software{Cole_Python_Tight_Binding_2025,
  author       = {Coh, Sinisa and
                  Vanderbilt, David and
                  Cole, Trey},
  title        = {Python Tight Binding (PythTB)},
  month        = nov,
  year         = 2025,
  publisher    = {Zenodo},
  version      = {2.0.0},
  doi          = {10.5281/zenodo.17595433},
  url          = {https://doi.org/10.5281/zenodo.17595433},
  note         = {available at https://doi.org/10.5281/zenodo.17595433},
  swhid        = {swh:1:dir:cba463f3222286f69cd57f382988f9e7514487ab
                   ;origin=https://doi.org/10.5281/zenodo.12721315;vi
                   sit=swh:1:snp:ee948998ca052bf922e18cfb7ecc21745514
                   bd57;anchor=swh:1:rel:80617f73ac2f572b590e0a8026ed
                   4acd9fb6a46f;path=pythtb-2.0.0
                  },
}

@article{soluyanov2011,
  title = {Wannier representation of ${\mathbb{Z}}_{2}$ topological insulators},
  author = {Soluyanov, Alexey A. and Vanderbilt, David},
  journal = {Phys. Rev. B},
  volume = {83},
  issue = {3},
  pages = {035108},
  numpages = {11},
  year = {2011},
  month = {Jan},
  publisher = {American Physical Society},
  doi = {10.1103/PhysRevB.83.035108},
  url = {https://link.aps.org/doi/10.1103/PhysRevB.83.035108}
}

@article{marzari2012,
  title = {Maximally localized Wannier functions: Theory and applications},
  author = {Marzari, Nicola and Mostofi, Arash A. and Yates, Jonathan R. and Souza, Ivo and Vanderbilt, David},
  journal = {Rev. Mod. Phys.},
  volume = {84},
  issue = {4},
  pages = {1419--1475},
  numpages = {0},
  year = {2012},
  month = {Oct},
  publisher = {American Physical Society},
  doi = {10.1103/RevModPhys.84.1419},
  url = {https://link.aps.org/doi/10.1103/RevModPhys.84.1419}
}

@article{ssh1979,
  title = {Solitons in Polyacetylene},
  author = {Su, W. P. and Schrieffer, J. R. and Heeger, A. J.},
  journal = {Phys. Rev. Lett.},
  volume = {42},
  issue = {25},
  pages = {1698--1701},
  numpages = {0},
  year = {1979},
  month = {Jun},
  publisher = {American Physical Society},
  doi = {10.1103/PhysRevLett.42.1698},
  url = {https://link.aps.org/doi/10.1103/PhysRevLett.42.1698}
}

\end{document}